\documentclass[twocolumn]{aa}
\usepackage{graphicx}
\usepackage{txfonts}
\newcommand{\mtot}{\relax \ifmmode M_{\rm tot}\else $M_{\rm tot}$\fi}
\newcommand{\Reff}{\relax \ifmmode R_{\rm e}\else $R_{\rm e}$\fi}
\newcommand{\SBe}{\relax \ifmmode \langle SB_{\rm e}\rangle \else $\langle SB_{\rm e}\rangle$\fi}
\newcommand{\mB}{\relax \ifmmode M_{\rm B}\else $M_{\rm B}$\fi}
\newcommand{\ReB}{\relax \ifmmode R_{\rm e,B}\else $R_{\rm e,B}$\fi}
\newcommand{\mueB}{\relax \ifmmode \mu_{\rm e,B} \else $\mu_{\rm e,B}$\fi}
\newcommand{\mD}{\relax \ifmmode M_{\rm D}\else $M_{\rm D}$\fi}
\newcommand{\muo}{\relax \ifmmode \mu_{\rm 0}\else $\mu_{\rm 0}$\fi}
\newcommand{\rd}{\relax \ifmmode h\else $h$\fi}
\newcommand{\db}{\relax \ifmmode D/B\else $D/B$\fi}
\newcommand{\nb}{\relax \ifmmode n\else $n$\fi}
\newcommand{\inc}{\relax \ifmmode i\else $h$\fi}
\newcommand{\ellip}{\relax \ifmmode \epsilon\else $\epsilon$\fi}

\newcommand{\magarc}{mag arcsec$^{-2}$}

\newcommand{\vmax}{\relax \ifmmode V_{\rm max}\else $V_{\rm max}$\fi}

\def\kms{\relax \ifmmode {\,\rm km\,s}^{-1}\else \,km\,s$^{-1}$\fi}
\def\ks{\relax \ifmmode  K_{\rm s}\else $K_{\rm s}$\fi}
\def\ha{\relax \ifmmode {\rm H}\alpha\else H$\alpha$\fi}
\def\hb{\relax \ifmmode {\rm H}\beta\else H$\beta$\fi}
\def\hi{\relax \ifmmode {\rm H\,{\sc i}}\else H\,{\sc i}\fi}
\def\hii{\relax \ifmmode {\rm H\,{\sc ii}}\else H\,{\sc ii}\fi}
\def\h2{\relax \ifmmode {\rm H}_2\else H$_2$\fi}
\def\lha{\relax \ifmmode L_{{\rm H}\alpha}\else $L_{{\rm H}\alpha}$\fi}
\def\shi{\relax \ifmmode \sigma_{{\rm HI}}\else $\sigma_{\rm HI}$\fi}
\def\sh2{\relax \ifmmode \sigma_{{\rm H}_2}\else $\sigma_{{\rm H}_2}$\fi}
\def\degr{\hbox{$^\circ$}}
\def\arcmin{\hbox{$^\prime$}}
\def\arcsec{\hbox{$^{\prime\prime}$}}
\def\deg{\hbox{$^\circ$}}
\def\min{\hbox{$^\prime$}}
\def\sec{\hbox{$^{\prime\prime}$}}
\def\fdg{\hbox{$.\!\!^\circ$}}
\def\fs{\hbox{$.\!\!^{\rm s}$}}
\def\farcm{\hbox{$.\mkern-4mu^\prime$}}
\def\farcs{\hbox{$.\!\!^{\prime\prime}$}}
\def\degd#1.#2{ #1\fdg#2 }                 
\def\mind#1.#2{ #1\farcm#2 }               
\def\secd#1.#2{ #1\farcs#2 }               
\def\hhh{\ifmmode {\rm ^h}              
         \else {${\rm ^h}$}
         \fi}
\def\sss{\ifmmode {\rm ^s}              
         \else {${\rm ^s}$}
         \fi}
\def\hms#1h#2m#3s{                      
                  \relax
                  \ifmmode #1^{\rm h}\,#2^{\rm m}\,#3^{\rm s}
                  \else \hbox{$#1^{\rm h}\,#2^{\rm m}\,#3^{\rm s}$}
                  \fi
                 }
\def\dms#1d#2m#3s{                      
                  \relax
                  #1\degr\,#2\arcmin\,#3\arcsec 
                 }
\def\hmsd#1h#2m#3.#4s{                  
                      \relax
                      \ifmmode #1^{\rm h}\,#2^{\rm m}\,#3\fs#4
                      \else \hbox{$#1^{\rm h}\,#2^{\rm m}\,#3\fs#4$}
                      \fi
                     }
\def\dmsd#1d#2m#3.#4s{                  
                      \relax
                      #1\degr\,#2\arcmin\,#3\farcs#4
                     }
\def\mag{\relax                          
        \ifmmode ^{\rm m}
        \else $^{\rm m}$
        \fi
       }
\def\magd#1.#2{                          
              \relax
              \ifmmode #1^{\rm m}
                       \hskip-0.55em.\hskip0.22em#2
              \else \hbox{#1$^{\rm m}
                    \hskip-0.55em.\hskip0.22em$#2}
              \fi
             }
\input psfig.sty
\begin{document}
\title{Massive star formation in the central regions of spiral 
galaxies}
\author{J.~H.~Knapen\inst{1}
\and L.~M.~Mazzuca\inst{2}
\and T.~B\"oker\inst{3}
\and I.~Shlosman\inst{4}
\and L.~Colina\inst{5}
\and F.~Combes\inst{6}
\and D.~J.~Axon\inst{1,7}
}

\offprints{J. H. Knapen}  

\institute{Centre for Astrophysics Research,
University of Hertfordshire, Hatfield, Herts AL10 9AB, U.K.\\
\email{j.knapen@star.herts.ac.uk} 
\and NASA Goddard Space Flight Center, Code 441, Greenbelt, MD 20771, 
USA
\and ESA/ESTEC, Keplerlaan 1, 2200 AG Noordwijk, Netherlands
\and Department of Physics and Astronomy,
University of Kentucky, Lexington, KY 40506-0055, USA
\and Instituto de Estructura de la Materia, CSIC, Serrano 119, 
E-28006 Madrid, Spain
\and Observatoire de Paris, 61 Av. de l'Observatoire. F-75\,014, Paris, 
France
\and Department of Physics, Rochester Institute of Technology, 84 
Lomb Memorial Drive, Rochester, NY, 14623, USA
}

\date{Received ; accepted}

\abstract{

{\it Context} The morphology of massive star formation in the central
regions of galaxies is an important tracer of the dynamical processes
that govern the evolution of disk, bulge, and nuclear activity.\\
{\it Aims} We present optical imaging of the central regions of a
sample of 73 spiral galaxies in the \ha\ line and in optical broad
bands, and derive information on the morphology of massive star
formation.\\
{\it Methods} We obtained images with the William Herschel Telescope,
mostly at a spatial resolution of below one second of arc. For most
galaxies, no \ha\ imaging is available in the literature. We outline
the observing and data reduction procedures, list basic properties,
and present the $I$-band and continuum-subtracted \ha\ images. We
classify the morphology of the nuclear and circumnuclear \ha\ emission
and explore trends with host galaxy parameters.\\
{\it Results} We confirm that late-type galaxies have a patchy
circumnuclear appearance in \ha, and that nuclear rings occur
primarily in spiral types Sa-Sbc. We identify a number of previously
unknown nuclear rings, and confirm that nuclear rings are
predominantly hosted by barred galaxies. \\
{\it Conclusions} Other than in stimulating nuclear rings, bars do not
influence the relative strength of the nuclear \ha\ peak, nor the
circumnuclear \ha\ morphology. Even though our selection criteria led
to an over-abundance of galaxies with close massive companions, we do
not find {\it any} significant influence of the presence or absence of
a close companion on the relative strength of the nuclear \ha\ peak,
nor on the \ha\ morphology around the nucleus.


}

\maketitle

%


\section{Introduction}

Enhanced nuclear activity in disk galaxies, in starburst or AGN form,
appears to be an integral part of their evolution. Both forms of
activity have been observed to co-exist (e.g., Heckman et al. 1997)
and are a clear manifestation of the symbiotic evolution of galactic
centres and their host galaxies.  The observed tight correlation
between the masses of the central black holes and the velocity
dispersions in the surrounding bulges (e.g., review by Ferrarese \&
Ford 2005) provides the most direct evidence for this evolution and
yields important clues on the dynamics, structure, and evolution of
galaxies.

To initiate and to maintain the AGN or nuclear starburst activity, gas
inflow must be stimulated from the disk to the central regions --- a
process which must be accompanied by a substantial loss of angular
momentum in the gas. Theoretically, this leads to the suggestion that
gravitational torques acting through galactic bars or galaxy
interactions are involved. Due to the asymmetric nature of their mass
distribution, they can facilitate the loss of angular momentum in
inflowing material (e.g., Shlosman, Begelman \& Frank 1990;
Athanassoula 1994; Combes 2001). Observationally, statistical links
between bars and interactions on the one hand and the occurrence of
starburst and AGN activity on the other are clear in certain
circumstances (e.g., extreme starbursts and interactions), barely
significant in others (e.g., low-$L$ AGN and bars; low-$L$ starbursts
and interactions), and absent in the case of low-$L$ AGN and
interactions (e.g., review by Knapen 2004).

Massive SF can be convincingly traced by the accompanying \ha\
emission and is very easily observed with standard telescopes and
cameras (Kennicutt 1998).  \ha\ is mainly produced in
the \hii\ regions surrounding massive B and O stars, although shocks
and non-stellar activity can also lead to \ha\ emission. In the images
of the 73 galaxies analysed here, we study the morphology of the \ha\
emission in the circumnuclear, two kpc radius regions, as well as from
the nucleus {\it per se}. The circumnuclear area as chosen is large
enough to incorporate most nuclear rings. This kind of circumnuclear,
low-$L$ starbursts are found in around one fifth of spiral galaxies
(Knapen 2005, hereafter K05), and characterise the dynamics of the
host galaxy and its stellar bar (e.g., Buta \& Combes 1996; K05).

We present our galaxy sample in Sect.~2, and describe the observations
and data reduction procedures in Sect.~3. The nuclear and
circumnuclear \ha\ morphology is analysed in Sections~4 and 5, and
relations to host galaxy properties, such as the presence of bars and
nuclear activity, the morphological type, and the presence of
companion galaxies, is discussed in Sect.~6. Section~7 lists our main
conclusions. In a subsequent Paper~II (L.~M.~Mazzuca et al., in
preparation) we will further investigate the morphology and stellar
ages in the 22 nuclear rings which we have identified in the course of
this study.


\section{Sample selection and  parameters}

For this study, we selected galaxies with some prior evidence for \ha\
structure in their central regions, either from the literature or from
our own unpublished work. Since one of the aims of the current study
is to identify nuclear rings, we have included a number of galaxies with
known nuclear or inner rings, some of which have also been described  in the
works by Pogge (1989a, b), Buta \& Crocker (1993), or K05. The
observed sample is therefore not complete and any results must be
interpreted with the appropriate care. For instance, the sample
selection procedure will not allow a determination of the fraction of,
say, nuclear rings in spiral galaxies.

The final sample, as observed by us with the William Herschel
Telescope (WHT), consists of 73 galaxies spanning a range
in many basic parameters, for which we obtained \ha, $B$ and $I$
imaging (see next Section for details). The galaxy sample is presented
in Table~\ref{sampletab}, which also lists a number of important
observational parameters as obtained from the literature. From the RC3
(de Vaucouleurs et al. 1991) we obtained the morphological type
(column~2 in Table~\ref{sampletab}), the apparent major isophotal
diameter measured at or reduced to surface brightness level
$\mu_B=25.0$\,$B$-\magarc, $D_{25}$ (in arcmin, col.~4), and the
inclination and major axis position angle (in degrees, col.~5 and
6). From the NASA/IPAC extragalactic database (NED) we obtained a
descriptor of the nuclear activity of the galaxy, for which we will
distinguish between the four main categories Seyfert, LINER, starburst
or H{\sc ii}, and none (col.~2). The recession velocity (in
km\,s$^{-1}$, col.~7) was obtained from the RC3 for most galaxies, but
from the NED for those where the RC3 does not list a value.

The distance $D$ to a galaxy (in Mpc, col.~8) was taken from the
Nearby Galaxies Catalog (Tully 1988) whenever a value was given there;
if not, we derived it from the recession velocity assuming a Hubble
constant of $H_0=75$\,km\,s$^{-1}$\,Mpc$^{-1}$. We used the distance
to derive the image scale, in parsec per arcsec (col.~10).  The
absolute blue magnitude ($M_B$, col.~11) was taken from Tully (1988),
or derived from the distance where a galaxy is not included in the
Tully (1988) catalogue.  As an indicator of bar strength, we list the
gravitational bar torque parameter $Q_{\rm b}$ (col.~12), as obtained
from various papers (see caption to Table~\ref{sampletab} for details
and Buta \& Block 2001 and Block et al. 2004 for a description of
$Q_{\rm b}$).

Finally, we list in Table~\ref{sampletab} (col. 16) whether a galaxy
has a close companion or not. The criterion for classifying a sample
galaxy as having such a companion is that it must have at least one
neighbouring galaxy within a radius of five times the diameter of the
galaxy under consideration, or $r_{\rm comp}<5\times D_{25}$, where
$D_{25}$ is listed in column~2 of Table~\ref{sampletab}. Any companion
must also have a recession velocity within a range of
$\pm200$~km\,s$^{-1}$ of the sample galaxy, and be not more than three
magnitudes fainter. These criteria have evolved from those used by
Schmitt (2001), Laine et al. (2002) and K05, and ensure that companion
galaxies are not only nearby, but also massive enough to have a
gravitational impact on the sample galaxy. The search for qualifying
neighbouring galaxies was performed using the ``near name'' search
option in the NED. 

We find that 26 of our 73 galaxies have a close companion (36\%). This
is a rather high number, no doubt influenced by the fact that we have
been selecting galaxies with previous evidence for ``interesting''
structure in their central regions, and also by the fact that our
sample spans a considerable range in distance. In comparison, only 48
(15\%) out of the 327 galaxies which comprise the statistically
complete sample of local galaxies in the HaGS (\ha\ Galaxy Survey,
James et al. 2004) show a close companion under the same criteria as
used here (J.~H.~Knapen \& P.~A.~James, in preparation).

\begin{table*}
\centering
\begin{tabular}{lccccccccccccccc}
\hline
NGC & Type & Activity & $D_{25}$ & $i$ & PA & $v$ & $D$ & Ref & scale & $M_B$ & $Q_{\rm b}$ & Ref. & \multicolumn{2}{c}{Morph.} & Comp.\\
& & (NED) & (\min) & (\deg) & (\deg) & (km\,s$^{-1}$) & (Mpc) & ($D$) & (pc/\sec) & & & ($Q_{\rm b}$) & N & CN & \\
\hline
~~128 & .L...P/ &  & 3.0 & 72 & 1 & 4241 & 56.6 & 1 & 274.1 &           -20.99 &  &  &         N & P & Y\\
~~157 & .SXT4.. &  & 4.2 & 50 & 40 & 1668 & 20.9 & 2 & 101.3 &          -20.06 & 0.33 & 8 &    W & P & N\\
~~210 & .SXS3.. &  & 5.0 & 49 & 160 & 1634 & 20.3 & 2 & 98.4 &          -20.06 & 0.05 & 8 &    N & P & Y\\
~~255 & .SXT4.. &  & 3.0 & 34 & 15 & 1600 & 20.0 & 2 & 97.0 &           -19.26 &  &  &         N & P & N\\
~~278 & .SXT3.. &  & 2.1 & 17 &  & 641 & 11.8 & 2 & 57.2 &              -19.62 & 0.05 & 8 &    N & R & N\\
~~470 & .SAT3.. & HII & 2.8 & 52 & 155 & 2374 & 30.5 & 2 & 147.9 &      -20.05 &  &  &         N & P & Y\\
~~473 & .SXR0*. &  & 1.7 & 51 & 153 & 2133 & 29.8 & 2 & 144.5 &         -19.77 &  &  &         W & R & N\\
~~488 & .SAR3.. &  & 5.2 & 42 & 15 & 2269 & 29.3 & 2 & 142.1 &          -21.36 & 0.03 & 8 &    S & N & N\\
~~613 & .SBT4.. & Sy & 5.5 & 41 & 120 & 1475 & 17.5 & 2 & 84.8 &        -20.53 & 0.30 & 8 &    S & R & N\\
~~628 & .SAS5.. &  & 10.5 & 24 & 25 & 656 & 9.7 & 2 & 47.0 &            -20.32 & 0.02 & 4 &    W & P & N\\
~~772 & .SAS3.. &  & 7.2 & 54 & 130 & 2458 & 32.6 & 2 & 158.0 &         -21.80 & 0.1 & 3 &     S & P & Y\\
~~788 & .SAS0*. & Sy2 & 1.9 & 41 & 75 & 4078 & 54.4 & 1 & 263.6 &       -20.68 &  &  &         S & N & N\\
~~864 & .SXT5.. &  & 4.7 & 41 & 20 & 1560 & 20.0 & 2 & 97.0 &           -20.20 & 0.32 & 8 &    S & P & N\\
~~922 & .SBS6P. &  & 1.9 & 36 &  & 3092 & 41.2 & 1 & 199.9 &            -20.63 &  &  &         N & P & Y\\
~~925 & .SXS7.. & HII & 10.5 & 56 & 102 & 553 & 9.4 & 2 & 45.6 &        -19.66 &  &  &         N & P & N\\
1042 & .SXT6.. &  & 4.7 & 39 & 15 & 1373 & 16.7 & 2 & 81.0 &            -19.91 & 0.04 & 8 &    W & P & Y\\
1068 & RSAT3.. & Sy1 Sy2 & 7.1 & 32 & 70 & 1137 & 14.4 & 2 & 69.8 &     -21.39 & 0.16 & 3,7 &  S & P & Y\\
1073 & .SBT5.. &  & 4.9 & 24 & 15 & 1211 & 15.2 & 2 & 73.7 &            -19.44 & 0.56 & 8 &    N & P & N\\
1079 & RSXT0P. &  & 3.5 & 52 & 87 & 1447 & 16.9 & 2 & 81.9 &            -18.83 &  &  &         S & P & N\\
1084 & .SAS5.. &  & 3.2 & 56 & 115 & 1406 & 17.1 & 2 & 82.9 &           -20.26 & 0.04 & 8 &    S & P & N\\
1087 & .SXT5.. &  & 3.7 & 53 & 5 & 1519 & 19.0 & 2 & 92.1 &             -20.21 & 0.43 & 8 &    S & P & N\\
1140 & .IB.9P* & HII Sy2 & 1.7 & 57 & 6 & 1509 & 18.2 & 2 & 88.2 &      -18.77 &  &  &         S & P & N\\
1232 & .SXT5.. &  & 7.4 & 29 & 108 & 1682 & 20.0 & 2 & 97.0 &           -21.11 & 0.21 & 7 &    N & N & N\\
1241 & .SBT3.. & Sy2 & 2.8 & 53 & 145 & 4030 & 26.6 & 2 & 129.0 &       -19.83 & 0.18 & 8 &    S & P & Y\\
1300 & .SBT4.. &  & 6.2 & 49 & 106 & 1568 & 18.8 & 2 & 91.1 &           -20.42 & 0.52 & 8 &    W & R & Y\\
1302 & RSBR0.. &  & 3.9 & 17 &  & 1703 & 20.0 & 2 & 97.0 &              -20.11 & 0.06 & 8 &    S & N & N\\
1343 & .SXS3*P &  & 2.6 & 51 & 80 & 2215 & 29.5 & 1 & 143.2 &           -18.85 &  &  &         W & R & N\\
1398 & PSBR2.. & Sy & 7.1 & 41 & 100 & 1407 & 16.1 & 2 & 78.1 &         -20.57 & 0.2 & 3,7 &   S & N & N\\
1530 & .SBT3.. &  & 4.6 & 58 &  & 2461 & 36.6 & 2 & 177.4 &             -21.32 & 0.71 & 6 &    N & R & N\\
1637 & .SXT5.. &  & 4.0 & 36 & 15 & 717 & 8.9 & 2 & 43.1 &              -18.33 & 0.19 & 8 &    S & P & N\\
3982 & .SXR3*. &  & 2.3 & 29 &  & 1109 & 17.0 & 2 & 82.4 &              -18.65 &  &  &         S & P & N\\
4303 & .SXT4.. & HII Sy2 & 6.5 & 27 &  & 1569 & 15.2 & 2 & 73.7 &       -20.71 & 0.08 & 8 &    S & R & N\\
4314 & .SBT1.. & LINER & 4.2 & 27 &  & 963 & 9.7 & 2 & 47.0 &           -18.65 & 0.44 & 8 &    S & R & Y\\
4321 & .SXS4.. & LINER HII & 7.4 & 32 & 30 & 1586 & 16.8 & 2 & 81.4 &   -21.13 & 0.18 & 3,4,5,7&S& R & Y\\
5248 & .SXT4.. & Sy2 HII & 6.2 & 44 & 110 & 1153 & 22.7 & 2 & 110.1 &   -21.07 & 0.06 & 8 &    S & R & N\\
5383 & PSBT3*P & Sbrst & 3.2 & 32 & 85 & 2250 & 37.8 & 2 & 183.3 &      -20.96 &  &  &         N & P & N\\
5430 & .SBS3.. & HII Sbrst & 2.2 & 58 & 0 & 2961 & 39.5 & 1 & 191.4 &   -20.11 &  &  &         N & P & N\\
5457 & .SXT6.. &  & 28.8 & 21 &  & 241 & 5.4 & 2 & 26.2 &               -20.45 & 0.125 & 3,5 & N & P & Y\\
5701 & RSBT0.. & LINER & 4.3 & 17 &  & 1506 & 26.1 & 2 & 126.5 &        -20.35 & 0.14 & 8 &    S & N & N\\
5728 & .SXR1*. & Sy2 & 3.1 & 55 & 0 & 2788 & 42.2 & 2 & 204.6 &         -21.67 &  &  &         S & R & N\\
5730 & .I..9*. &  & 1.8 & 78 & 88 & 2533 & 33.8 & 1 & 163.7 &           -18.01 &  &  &         N & P & Y\\
5850 & .SBR3.. &  & 4.3 & 29 & 140 & 2556 & 28.5 & 2 & 138.2 &          -20.69 & 0.31 & 8 &    S & P & Y\\
5905 & .SBR3.. & Sy1 & 4.0 & 49 & 135 & 3390 & 45.2 & 1 & 219.1 &       -20.69 & 0.43 & 5 &    N & R & Y\\
5921 & .SBR4.. & LINER & 4.9 & 36 & 130 & 1480 & 25.2 & 2 & 122.2 &     -20.67 & 0.26 & 8 &    S & P & N\\
5945 & .SBT2.. &  & 2.1 & 21 & 105 & 5516 & 73.6 & 1 & 356.6 &          -20.63 &  &  &         S & R & Y\\
5953 & .SA.1*P & LINER Sy2 & 1.6 & 34 & 169 & 1965 & 33.0 & 2 & 160.0 & -19.59 &  &  &         S & R & Y\\
\hline
\end{tabular}
\caption{Global parameters of the galaxies in the observed sample,
obtained from the RC3 unless otherwise indicated. Galaxies are listed
in order of increasing RA. Tabulated are the identification number of
the sample galaxies (all are NGC numbers except IC~1438 (col.~1); the
morphological type (col.~2); nuclear activity class (from NED;
col.~3); $D_{25}$, the apparent major isophotal diameter measured at
or reduced to surface brightness level $\mu_B=25.0$\,$B$-\magarc
(col.~4); inclination $i$ as derived from the ratio of the major to
the minor isophotal diameter (col.~5); position angle PA of the disk
(col.~6); mean heliocentric radial velocity $v$, in a few cases
obtained from NED (col.~7); distance $D$ in Mpc (col.~8); reference
for $D$, where 1 is $v$ from col.~7 and assuming a Hubble constant of
$H_0=75$km\,s$^{-1}$\,Mpc$^{-1}$, 2 is the Nearby Galaxies Catalog
(Tully 1988; col.~9); image scale in parsec per arcsec, as derived
from the distance (col.~10); absolute blue magnitude, taken from Tully
(1988) whenever possible, otherwise derived from $m_B$ (from the RC3)
and the distance (col.~11); gravitational bar torque $Q_{\rm b}$
(col.~12); reference for $Q_{\rm b}$ (col.~13), where 3 is Laurikainen
\& Salo (2002), 4 is Block et al. (2001), 5 is Buta \& Block (2001), 6
is Block et al. (2004), 7 is Laurikainen et al. (2004), 8 is Buta et
al. (2005). An average is given when more than one measure is
available in the literature, except where a value is available from
Block et al.  (2004) or Buta et al. (2005), which are given without
taking earlier determinations into account; morphological
classification of the nuclear (col.~14; where 'N' means no emission,
'W' weak, 'S' strong, and 'P' position; see Sect.~4.1) and
circumnuclear \ha\ emission (col.~15; 'D' means diffuse, 'P' patchy,
'N' none, 'R' ring; see Sect.~5.1), and whether a galaxy is classed as
having a close companion ('Y') or not ('N'; col.~16; see Sect.~2).}
\label{sampletab}
\end{table*}

\setcounter{table}{0}
\begin{table*}
\centering
\begin{tabular}{lccccccccccccccc}
\hline
NGC & Type & Activity & $D_{25}$ & $i$ & PA & $v$ & $D$ & Ref & scale & $M_B$ & $Q_{\rm b}$ & Ref. & \multicolumn{2}{c}{Morph.} & Int.\\
& & (NED) & (\min) & (\deg) & (\deg) & (km\,s$^{-1}$) & (Mpc) & ($D$) & (pc/\sec) & & & ($Q_{\rm b}$) & N & CN & \\
\hline
5970 & .SBR5.. & LINER HII & 2.9 & 47 & 88 & 1963 & 31.6 & 2 & 153.2 &  -20.55 &  &  &         N & P & Y\\
5982 & .E.3... &  & 2.6 & 41 & 110 & 3017 & 40.2 & 1 & 195.0 &          -20.98 &  &  &         S & N & Y\\
6217 & RSBT4.. & Sy2 & 3.0 & 34 &  & 1362 & 23.9 & 2 & 115.9 &          -20.19 & 0.39 & 3 &    S & P & N\\
6384 & .SXR4.. & LINER & 6.2 & 49 & 30 & 1663 & 26.6 & 2 & 129.0 &      -21.31 & 0.14 & 8 &    S & N & N\\
6412 & .SAS5.. &  & 2.5 & 29 &  & 1324 & 23.5 & 2 & 113.9 &             -19.69 &  &  &         N & P & N\\
6503 & .SAS6.. & LINER HII & 7.1 & 70 & 123 & 44 & 6.1 & 2 & 29.6 &     -18.64 &  &  &         W & R & N\\
6574 & .SXT4*. & Sy & 1.4 & 39 & 160 & 2282 & 35.0 & 2 & 169.7 &        -20.76 &  &  &         W & P & N\\
6814 & .SXT4.. & Sy1.5 & 3.0 & 21 &  & 1563 & 22.8 & 2 & 110.5 &        -20.41 &  &  &         S & P & N\\
6907 & .SBS4.. &  & 3.3 & 36 & 46 & 3161 & 42.2 & 1 & 204.3 &           -21.22 & 0.07 & 8 &    S & P & N\\
6946 & .SXT6.. & HII & 11.5 & 32 &  & 52 & 5.5 & 2 & 26.7 &             -20.78 & 0.04 & 4 &    S & P & N\\
6951 & .SXT4.. & LINER Sy2 & 3.9 & 34 & 170 & 1426 & 24.1 & 2 & 116.8 & -20.73 & 0.39 & 6 &    S & R & N\\
7130 & .S..1P. & LINER Sy2 & 1.5 & 24 &  & 4842 & 64.6 & 1 & 313.0 &    -21.13 &  &  &         S & P & N\\
7217 & RSAR2.. & LINER Sy & 3.9 & 34 & 95 & 946 & 16.0 & 2 & 77.6 &     -20.38 & 0.03 & 8 &    S & R & N\\
IC~1438 & PSXT1*. &  & 2.4 & 32 &  & 2616 & 33.8 & 2 & 163.9 &          -20.08 &  &  &         W & R & Y\\
7331 & .SAS3.. & LINER & 10.5 & 69 & 171 & 821 & 14.3 & 2 & 69.3 &      -21.10 & 0.12 & 3 &    S & P & Y\\
7469 & PSXT1.. & Sy1.2 & 1.5 & 44 & 125 & 4916 & 65.6 & 1 & 317.8 &     -21.08 &  &  &         S & R & Y\\
7479 & .SBS5.. & LINER Sy2 & 4.1 & 41 & 25 & 2378 & 32.4 & 2 & 157.1 &  -21.11 & 0.70 & 8 &    S & P & N\\
7550 & .LA.-.. & AGN & 1.4 & 29 &  & 5072 & 67.6 & 1 & 327.9 &          -20.99 &  &  &         W & N & Y\\
7570 & .SB.1.. &  & 1.5 & 54 & 30 & 4698 & 62.6 & 1 & 303.7 &           -20.09 &  &  &         N & R & N\\
7606 & .SAS3.. &  & 5.4 & 67 & 145 & 2233 & 28.9 & 2 & 140.1 &          -21.28 & 0.09 & 3 &    S & D & N\\
7672 & .S..3.. & Sy2 & 1.2 & 41 &  & 4010 & 53.5 & 1 & 259.2 &          -18.94 &  &  &         S & P & Y\\
7716 & .SXR3*. &  & 2.1 & 34 & 35 & 2571 & 33.7 & 2 & 163.4 &           -19.84 &  &  &         S & R & N\\
7723 & .SBR3.. &  & 3.5 & 47 & 35 & 1875 & 23.7 & 2 & 114.9 &           -20.29 & 0.32 & 8 &    S & P & N\\
7727 & .SXS1P. &  & 4.7 & 41 & 35 & 1814 & 23.3 & 2 & 113.0 &           -20.38 & 0.09 & 8 &    S & D & Y\\
7741 & .SBS6.. &  & 4.4 & 47 & 170 & 755 & 12.3 & 2 & 59.6 &            -18.76 & 0.74 & 8 &    P & P & N\\
7742 & .SAR3.. & LINER HII & 1.7 & 0 &  & 1653 & 22.2 & 2 & 107.6 &     -19.65 &  &  &         S & R & N\\
7769 & RSAT3.. & LINER HII & 1.7 & 17 &  & 4214 & 56.2 & 1 & 272.4 &    -20.93 &  &  &         S & P & Y\\
\hline
\end{tabular}
\caption{Continued.}
\end{table*}


\section{Observations and data reduction}

All the imaging presented in this paper has been obtained using the
Auxiliary Port (Aux Port) camera on the 4.2m WHT, operated by the
Isaac Newton Group in La Palma. The bulk of the imaging was obtained
during a total of four observing nights granted by the UK time
allocation panel (1999 September 16 and 17, and 2000 July 25 and 26),
with additional imaging obtained in service mode during a number of
nights in 1999 and 2000. We also used images of a few galaxies (e.g.,
NGC~4314 and NGC~4321) obtained by one of us (JHK) during earlier
observing runs with the WHT. The images of NGC~4321 have been
published before by Knapen et al. (1995a,b).

The Aux Port camera is a small optical camera without any re-imaging
optics located at a dedicated folded cassegrain focus of the WHT.  Its
small pixel scale (0.11 arcsec/pixel) combined with a 1024$\times$1024
pixel TEK CCD and round filters yields a circular field of view (FOV)
of about 1.8 arcmin in diameter. We used Harris $B$ and $I$ filters,
and one of the narrow-band filters 6570/55, 6594/44, 6607/50, 6626/44
or 6656/44, depending on the systemic velocity of the galaxy, where
the numbers denote $\lambda/\Delta\lambda$, or the central wavelength
of transmission and the width of the transmission curve, both in \AA.
We used typical exposure times of 1, 3, and 10 minutes in $I$ and $B$,
and \ha, respectively.  Atmospheric conditions were good in general,
and the resulting spatial resolution as measured from the reduced
images is around 0.8 arcsec in the \ha\ and $I$-band images, and
around 1.0~arcsec in $B$.

The data reduction process followed basic procedures for bias
subtraction and flat fielding using twilight sky flat fields.  Image
registration was done by measuring the positions of foreground stars
where available, and the centre of the galaxy in all cases. For the
subtraction of the continuum from the \ha\ images we used the $I$-band
images.  We determined the scaling factor by applying the procedure
described by B\"oker et al.  (1999) to all individual sets of \ha\ $+$
$I$ images. This method uses the fact that most pixels in any set of
two registered images of a galaxy will show continuum emission,
whereas only a minority show continuum plus line emission.  When
plotting all individual pixels, those tracing the continuum will
scatter along a narrow, well-defined band, the slope of which denotes
the scaling factor to be used for the continuum subtraction.  Pixels
containing \ha\ emission will end up above this band (assuming that
\ha\ intensity is plotted as the ordinate).  We refer the reader to
the papers by B\"oker et al.  (1999) and Knapen et al. (2004a) for a
more detailed description. Knapen et al. (2004a) discuss the
uncertainties introduced by using $I$-band imaging for continuum
subtraction, and conclude that the resulting errors in the
luminosities of H{\sc ii} regions are smaller than some 5\%, and that
the resulting \ha\ morphology is reliable.

\begin{figure*}
\psfig{figure=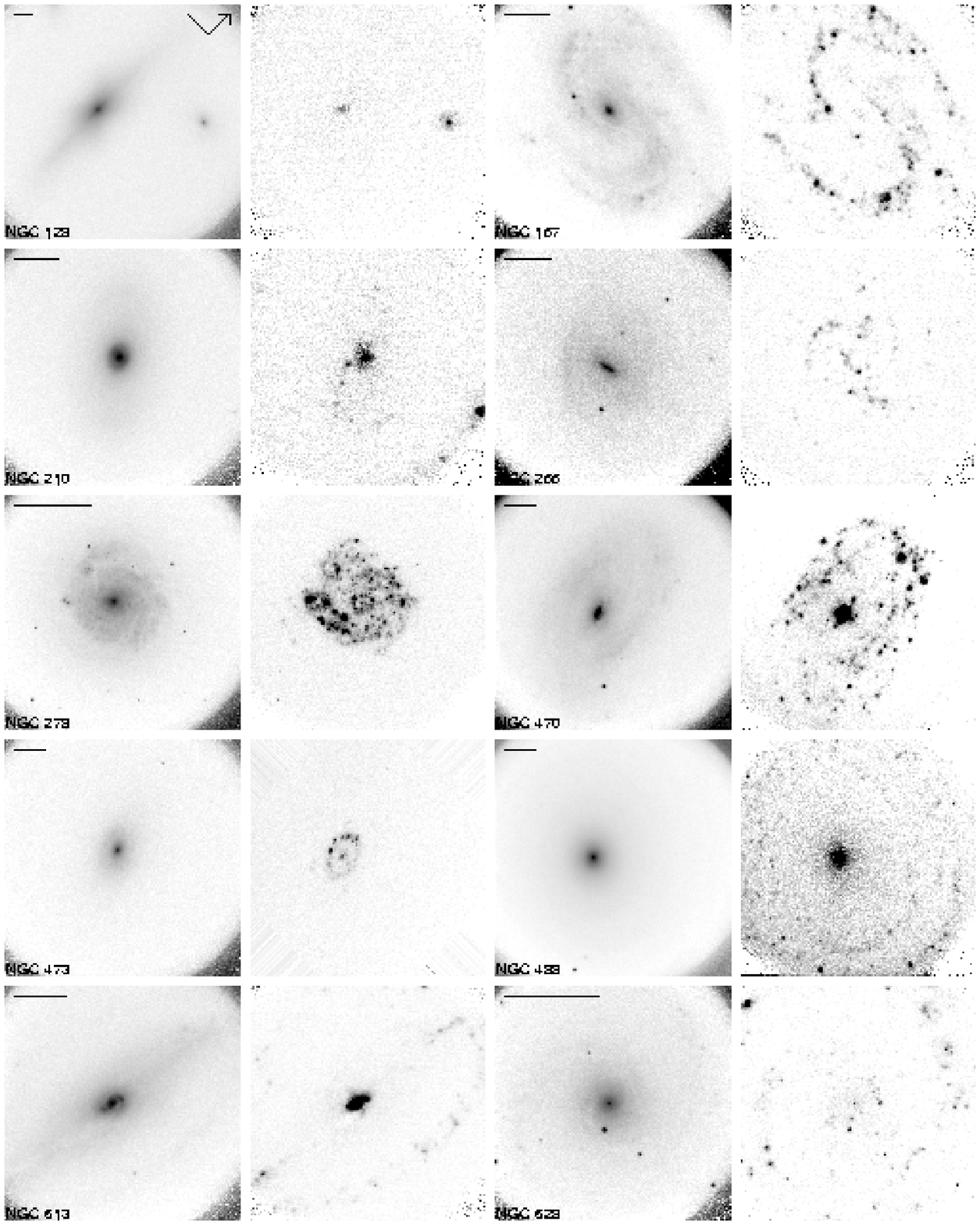,width=17cm}
\caption{***Figure consists of eight panels of thumbnail images, to be
published electronically only*** Grey-scale representation of the
$I$-band and continuum-subtracted \ha\ images of all sample galaxies,
with the former shown on the left in each pair. The orientation is
North up and East to the left, except for those cases where East and
North are indicated in the upper right corner of the $I$-band image,
and where the arrow points North. The scale is indicated for each
galaxy by the length of the black line in the top left corner of the
$I$-band image, which corresponds to 2~kpc.}
\label{images}
\end{figure*}

The $I$-band and continuum-subtracted \ha\ images of all sample
galaxies are shown in Fig.~1 (published electronically).  The $B$-band
images (not shown here) generally outline a morphology similar to that
shown in \ha, though less pronounced in showing the star forming
regions. 


\section{\ha\ morphology: Methodology and limitations}

\subsection{Nuclear \ha\ emission}

We have chosen to limit the classification of the nuclear \ha\
morphology to three categories: strong, weak, and none. This is mainly
because of the uncertainties in the continuum subtraction using
$I$-band images in the very centres of our galaxies, where
differential dust extinction can play a more prominent role than in
the disk (Knapen et al. 2004a). These classifications apply to the
central point source --- in practise the nuclear region of one seeing
element in size. The exact position of the nucleus in a galaxy was
determined from the $I$-band image. To qualify as a strong \ha\
source, a nuclear point source must be more luminous than any other
\ha-emitting region in the galaxy. To qualify as weak, a nuclear \ha\
peak must be present but stronger peaks are found outside the
nucleus. Obviously, our category ``none'' describes those cases where
no believable \ha\ emission can be observed from the nucleus. In one
case (that of NGC~7741) we could not pinpoint the location of the
nucleus (not even using the near-IR image from Knapen et al. 2003),
and this galaxy is thus not included in the following discussion.

\subsection{Circumnuclear \ha\ morphology}

For the purpose of this paper, we consider the morphology of the \ha\
emission in the circumnuclear region as that arising from outside the
nucleus and within a radius of 2.0~kpc from the nucleus of the
galaxy. The range in distances to our sample galaxies, from 5 to 74~Mpc
(Table~\ref{sampletab}), implies that this 2~kpc radius corresponds to a
range of linear sizes in our images, varying roughly from 77~arcsec to
6~arcsec. The \ha\ morphology was classified by visual inspection of the
images within the appropriate radial area independently by two of the
co-authors of this paper.

We used a minimal number of different morphological classes, following
the scheme introduced by K05. The main categories are, first,
``none'', where obviously no \ha\ emission is detected in the
circumnuclear region (there will usually be emission from the
nucleus). Second, ``patchy'', where individual and clearly delimited
patches of \ha\ emission are detected, tracing individual \hii\
regions, but not in an obvious ring pattern. The ``patchy'' class
includes galaxies where the circumnuclear \hii\ regions form a spiral
pattern, but because such a pattern is hard to classify simply and
unambiguously, we have not explicitly classified such cases as
spiral. Third, ``ring'', which denotes \ha\ emission organised into a
well-defined nuclear ring. In three galaxies, circumnuclear \ha\ is
detected which is not confined to individual patches, and these three
galaxies have been classified ``diffuse''. The classifications are
given in Table~\ref{sampletab}.

Of the 22 nuclear ring galaxies reported here, to the extent of our
knowledge five have not been reported as nuclear ring hosts in the
literature. These newly identified nuclear rings are those in
NGC~473, NGC~5905, NGC~5953, NGC~7570, and NGC~7716.

The classification scheme used here is rather coarse, but we believe
that it offers a good compromise between the need for classification
and the enormous variety encountered in the detailed \ha\ morphology
from galaxy to galaxy. This variety is no doubt partly due to the
nature of \ha\ emission - originating in regions of massive star
formation whose ionised gas emission will show a strong time
evolution.

We will not discuss the fractions of galaxies in our sample which show
patchy, ring-shaped, or no \ha\ emission, because the selection
criteria of our sample imply that any conclusions from such a
discussion would have no wider significance. This is in contrast to
the study of K05, where the selection of a statistically meaningful
sample leads, for instance, to the result that one of every five
spiral galaxies has a nuclear ring. What the current sample can be used
for, except of course for selecting sub-samples of specific interest,
or for referring to individual objects, is an analysis of how a
certain circumnuclear \ha\ morphology is distributed with respect to
host galaxy parameters such as morphological class or presence of a
bar, provided that reference is made to the distribution of such
parameters across the whole sample under consideration (next Section).


\section{\ha\ morphology and host galaxy parameters}

\subsection{Overall results}

\begin{table}
\centering
\begin{tabular}{lccccccc}
\hline
Emission & $N$ & $D$ & $d$ & $M_B$ & $Q_{\rm b}$ & \multicolumn{2}{c}{Companions?}\\
& & (\min) & (Mpc) & (mag) & & Y (\%) & N (\%)\\
\hline
\multicolumn{7}{c}{Overall sample}\\
       & 73 & 3.7 & 23.9 & -20.4 & 0.18 & 36$\pm$6\% & 64$\pm$6\%\\
\multicolumn{7}{c}{Nuclear emission}\\
Strong & 44 & 3.9 & 23.8 & -20.6 & 0.16 & 32$\pm$7\% & 68$\pm$7\%\\
Weak   & 10 & 3.4 & 24.5 & -20.1 & 0.19 & 40$\pm$15\% & 60$\pm$15\%\\
None   & 18 & 3.0 & 31.0 & -20.1 & 0.21 & 44$\pm$12\% & 56$\pm$12\%\\
\multicolumn{7}{c}{Circumnuclear emission}\\
Patchy & 39 & 3.5 & 22.8 & -20.2 & 0.19 & 36$\pm$8\% & 64$\pm$8\%\\
None   & 9  & 4.3 & 26.6 & -21.0 & 0.14 & 22$\pm$14\% & 78$\pm$14\%\\
Ring   & 22 & 3.5 & 26.5 & -20.4 & 0.30 & 36$\pm$10\% & 64$\pm$10\%\\
Diffuse& 3  & 4.7 & 28.9 & -21.0 & 0.09 & 67$\pm$27\% & 33$\pm$27\%\\
\hline
\end{tabular}
\caption{Median values of the host galaxies' diameter (in arcmin, from
the RC3; col.~3), distance (in Mpc, from Tully 1988; col.~4), absolute
magnitude (calculated using $m_B$ from the RC3 and the distance;
col.~5), and the gravitational bar strength $Q_{\rm b}$ (col.~6), for
the overall sample and the different classes of nuclear and
circumnuclear emission (col.~1; the number of galaxies in each
category is shown in col.~2). The last two columns show the fractions
of the different nuclear and circumnuclear \ha\ emission morphologies
among those galaxies which have close companions (col.~6) and those
which do not (col.~7), following the criteria set out in
Sect.~2. Uncertainties are 1~$\sigma$ as determined from Poisson
statistics.}
\label{overalltab}
\end{table}

The results on the overall distribution of the different \ha\
morphology classes are summarised in Table~\ref{overalltab}. Strong
nuclear emission is found in 44 of our 73 sample galaxies, while 10
are classified as having weak nuclear \ha\ emission, and 18 as having
no \ha\ emission at all. The circumnuclear \ha\ emission is patchy in
more than half of our galaxies (39 out of 73), and in the form of a
nuclear ring in 22. A further 9 galaxies show no
circumnuclear \ha\ emission at all, and 3 have diffuse emission.

Table~\ref{overalltab} also lists the median values of a number of key
host galaxy parameters for the overall sample and for the different
classes of nuclear and circumnuclear \ha\ emission. We have listed
diameter, distance, absolute blue magnitude, and the bar's gravitational
torque parameter $Q_{\rm b}$, or bar strength, for those galaxies
where this value is given in the literature. The only trend worth
noting for the nuclear \ha\ emission is that those galaxies without
\ha\ emission from the nucleus are, in the median, further away than
those with emission, which is easily understood as related to
sensitivity and detection. Those galaxies with strong \ha\ peaks in
the nucleus are slightly brighter than others, but this may be a
morphological type effect (see next section) and we will not discuss
it in more detail.

As for the circumnuclear \ha\ emission, Table~\ref{overalltab} shows
how the 12 galaxies without, or with diffuse, \ha\ emission are both
larger in diameter and fainter than the others. Those with patchy \ha\
emission are, in the median, at smaller distances from us. Using
$Q_{\rm b}$ as an indicator, the galaxies with nuclear rings appear to
have slightly stronger bars in the median, but a Kolmogorov-Smirnov
(KS) test\footnote{The Kolmogorov-Smirnov test is a statistical test
that describes the likelihood that two data sets have been drawn from
the same parent distribution: a probability $P<0.05$ is usually
interpreted as describing two statistically distinct populations.}
shows that this difference is not statistically significant.

We also list the percentages of galaxies in the different categories
which have nearby companions. Taking into account the numbers of
galaxies in the different categories, the presence or absence of a
close companion  influences {\it neither} the nuclear, {\it
nor} the circumnuclear \ha\ morphology. We will discuss this result,
and its implications, in more detail in Sect.~5.3.

\subsection{Morphological type of the host galaxy}

\begin{table*}
\centering
\begin{tabular}{lcccccccccccccccc}
\hline
Emission & $N$ & \multicolumn{11}{c}{Morphological type $T$} & 
\multicolumn{4}{c}{Spiral type}\\
 & & 0 & 1 & 2 & 3 & 4 & 5 & 6 & 7 & 8 & 9 & $-$ & SA & SX & SB & $-$\\
 & & S0 & Sa & Sab & Sb & Sbc & Sc & Scd & Sd & Sdm & Sm & - &  
\multicolumn{4}{c}{$N$ (\%)}\\
\hline
\multicolumn{17}{c}{Total sample}\\
 &       73 & 5 & 8 & 3 & 22& 14& 10& 6 & 1 & 0 & 2 & 2 & 16 (22) & 28 (38) & 24 (33) & 5 (7)\\
\multicolumn{17}{c}{Nuclear emission}\\
Strong & 44 & 4 & 6 & 3 & 13& 10& 5 & 1 & 0 & 0 & 1 & 1 & 11 (25) & 16 (36) & 14 (32) & 3 (7)\\
Weak   & 10 & 1 & 1 & 0 & 1 & 3 & 1 & 2 & 0 & 0 & 0 & 1 & 3 (30) & 6 (60) & 1 (10)  & 0 (0)\\
None   & 18 & 0 & 1 & 0 & 7 & 1 & 4 & 2 & 1 & 0 & 1 & 1 & 2 (11) & 6 (33) & 8 (44) & 2 (11)\\
Position& 1 &   &   &   &   &   &   & 1 &   &   &   &   &    &    &    &\\
\multicolumn{17}{c}{Circumnuclear emission}\\
Patchy & 39 & 1 & 1 & 0 &13 & 7 & 9 & 5 & 1 & 0 & 2 & 0 & 8 (21) & 14 (36) & 14 (36) & 3 (8)\\
None &    9 & 3 & 0 & 1 & 1 & 1 & 1 & 0 & 0 & 0 & 0 & 2 & 3 (33) & 2 (22) & 3 (33) & 1 (11)\\
Ring &   22 & 1 & 6 & 2 & 6 & 6 & 0 & 1 & 0 & 0 & 0 & 0 & 4$^1$ (18) & 11 (50) & 7 (32) & 0 (0)\\
Diffuse & 3 & 0 & 1 & 0 & 1 & 0 & 0 & 0 & 0 & 0 & 0 & 1 & 1 (33) & 1 (33) & 0 (0) & 1 (33)\\
\hline
\end{tabular}
\caption{Morphological and bar type distribution of the various types
of circumnuclear \ha\ emission.  For the whole sample and for the
nuclear and circumnuclear emission, col.~2 gives the total number of
galaxies in each category (as described in col.~1), with the number of
galaxies of each morphological type (from the RC3) in columns 3-13,
and bar type (from the RC3) in columns 14-17. In these four columns,
the numbers in brackets are the percentages of the total. Note $^1$:
see Sect.~5.5.}
\label{typetab}
\end{table*}

\setcounter{figure}{1}
\begin{figure}
\psfig{figure=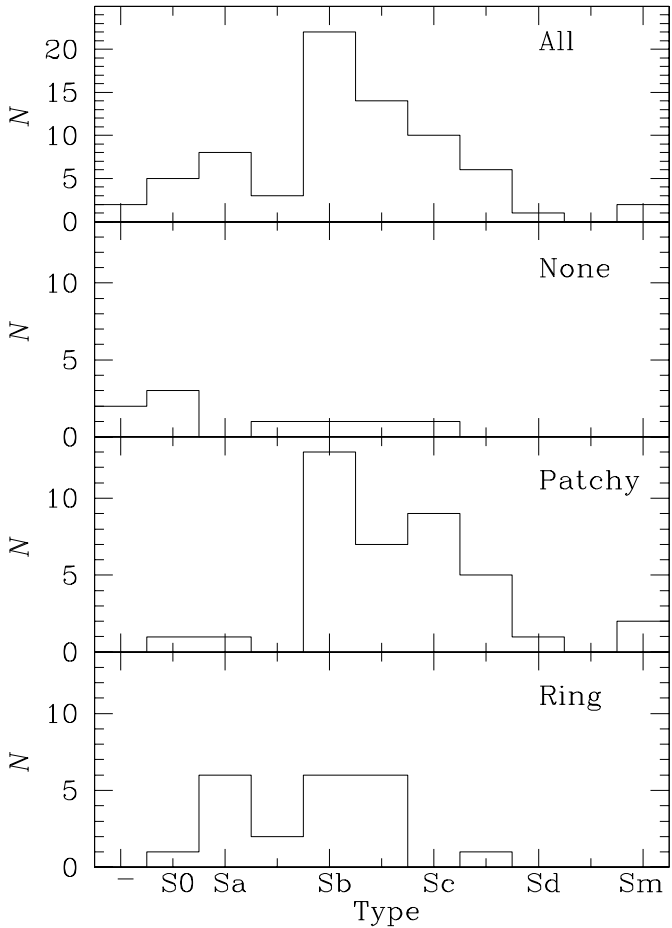,angle=0,width=9cm}
\caption{Set of histograms showing the distribution of circumnuclear
\ha\ emission morphology with host galaxy type, the latter obtained
from the RC3.}
\label{typecn}
\end{figure}

\begin{figure}
\psfig{figure=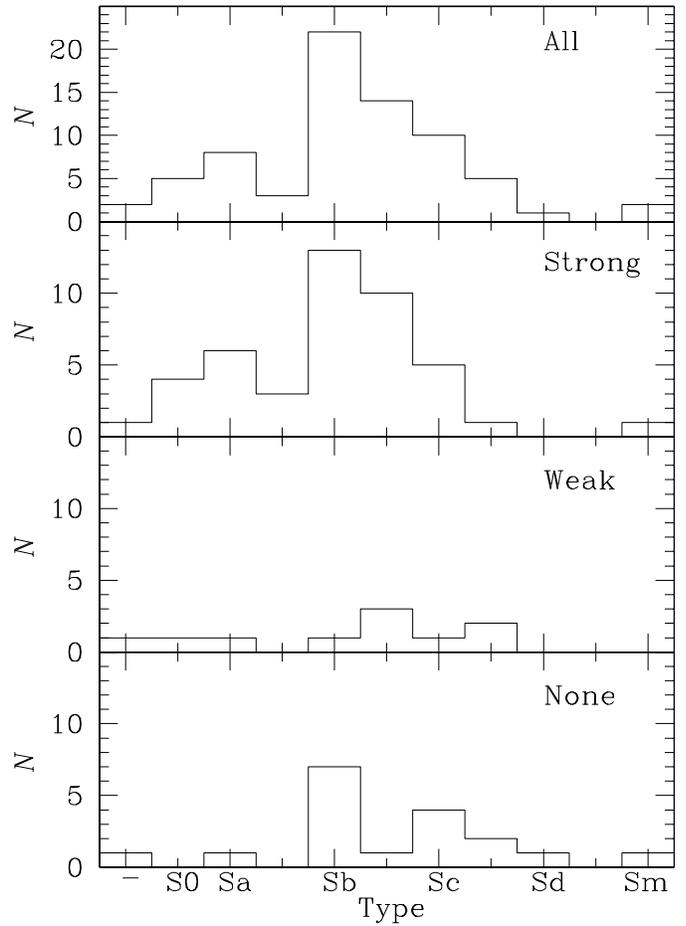,angle=0,width=9cm}
\caption{As Fig.~2, now for the  distribution of nuclear
\ha\ emission morphology.}
\label{typen}
\end{figure}


\subsubsection{$T$-type}

Table~\ref{typetab} lists the distribution of the various classes of
nuclear and circumnuclear morphology with spiral $T$-type, as obtained
from the morphological classifications listed in the RC3 (the
translation of $T$-type as used in the RC3 and more widely used
descriptors of the type Sa, Sb, etc., is given in the header to
Table~\ref{typetab}). These results are shown as histograms in
Figs.~2 and 3 for the circumnuclear and nuclear
\ha\ morphology, respectively. The results confirm those presented by
K05. The overall distribution of $T$-type for the sample shows a peak
near types Sb and Sbc ($T=$3 and 4, see Table~\ref{typetab}), a broad
distribution ranging from S0 to Sm, and a lack of galaxies of type
Sdm. For 2 galaxies, no $T$-type is given in the RC3. We reiterate
that no attempts were made at the sample selection stage to achieve a
specific type distribution.

Galaxies with patchy {\it circumnuclear} \ha\ morphology tend to be of
later types, and in fact there are only two galaxies of the 19 of type
later than Sbc which are {\it not} classified as patchy (an Sc with no
circumnuclear \ha, and an Scd with a ring). Ring-like morphology is
preferentially found in earlier types, peaking at Sb and Sbc
types. Galaxies with no circumnuclear \ha\ emission are also
preferentially of early type, confirming the findings of K05.

Not much can be deduced from the distribution of {\it nuclear} \ha\
morphology with morphological type of the host galaxy (Fig.~3) because
of the relatively low numbers of galaxies with weak, or no, nuclear
\ha\ emission, but the $T$-type of the host does not seem to correlate
with the presence of relative strength of the nuclear \ha\ emission.

\subsubsection{Bars}

The distribution of \ha\ morphology has been tabulated
(Table~\ref{typetab}) for the various bar classifiers used in the RC3:
SA, SX (often quoted as SAB), and SB, in Table~\ref{typetab}. Five
galaxies have merely been classified as 'S' in the RC3 and are not
considered here. The distribution between SA, SX, and SB classes seen
across the whole sample is closely mirrored in the sub-samples of
different nuclear and circumnuclear \ha\ morphology. The numbers in
the Table might appear to indicate a slight excess of SX galaxies
among those with weak nuclear \ha, and a slight lack of SA galaxies
among those with no nuclear emission. Poisson statistics show that in
either case the significance of this difference is at most
1$\sigma$. Considering the same number from the perspective of
non-barred vs. barred galaxies, we see that 69\%$\pm$12\% of the
non-barred (SA) vs. 58\%$\pm$7\% of the barred (SX$+$SB) galaxies has
``strong'' nuclear \ha\ emission, 19\%$\pm$10\% vs. 13\%$\pm$5\%
``weak'', and 13\%$\pm$8\% vs. 27\%$\pm$6\% ``none''. Again, none of
the differences are significant.

The bar distribution in the galaxies in the different classes of
circumnuclear emission mirror those in the overall sample closely (the
apparent deviations in the case of the three SA and two SX ``none''
galaxies are a 1$\sigma$ effect). Table~\ref{typetab} shows explicitly
how the percentages of barred and non-barred galaxies are rather
stable for the different classes of morphology. Calculating the
percentage of the different morphology classes for barred and
non-barred galaxies, we see the same effect, for instance,
50\%$\pm$13\% of non-barred (SA) galaxies have a patchy circumnuclear
\ha\ morphology, versus 54\%$\pm$7\% of barred galaxies
(SX$+$SB). These numbers are 19\%$\pm$10\% vs. 10\%$\pm$4\% for
``none''; and 25\%$\pm$11\% vs. 35\%$\pm$7\% for ``ring'' (but see
below).

The galaxies hosting circumnuclear rings in \ha\ only show a rather
moderate preference for SX and SB hosts, which at first sight may
appear to be at odds with the findings of K05, who confirmed the
standard picture that bars are almost exclusively responsible for the
existence of nuclear rings. In fact, K05 pointed out that of the 12
nuclear ring hosts considered, only two have been originally
classified as non-barred SA, but that each of those two has clear
evidence from near-IR imaging for the existence of a bar.  We will
discuss the four SA-type nuclear ring host galaxies in the current
sample in more detail in Sect.~5.5.

\subsection{Close companions}

\begin{table*}
\centering
\begin{tabular}{lcccccc|ccccc|ccc}
\hline
  & $N$ & $D$ & $d$ & $M_B$ & type & $Q_{\rm b}$ & 
\multicolumn{4}{c}{Circumnuclear \ha\ morph. ($N$(\%))} && 
\multicolumn{3}{c}{Nuclear \ha\ morph. ($N$(\%))}\\
  & & (\min) & (Mpc)& (mag) &&& Patchy & None & Ring & Diff. && Strong & 
Weak & None\\
\hline
Close companion     & 26 & 2.9 & 32.1 & $-$20.6 & 4.0 & 0.16 & 14 (54) & 2 (8) & 8 (31) & 2 (8) && 14 (54) & 4 (15) & 8 (31)\\
No close companion & 47 & 3.9 & 22.7 & $-$20.3 & 3.0 & 0.20 & 25 (53) & 7 (15) & 14 (30) & 1 (2) && 30 (64) & 6 (13) & 10 (21)\\
\hline
\end{tabular}
\caption{Median values of overall parameters of the interacting (see
Sect.~2 for criteria) and non-interacting sample galaxies (sample
sizes in col.~2), as well as the numbers and percentages (latter in
parenthesis) of those galaxies in the different circumnuclear
(cols.~8-11) and nuclear (cols.~12-14) \ha\ morphology
classes. Parameters are diameter (in arcmin, from the RC3; col.~3),
distance (in Mpc, from Tully 1988; col.~4), absolute magnitude
(calculated using $m_B$ from the RC3 and the distance; col.~5),
morphological type (from the RC3; col.~ 6), and the gravitational bar
strength $Q_{\rm b}$ (col.~7).}
\label{companion_props}
\end{table*}

In analysing the relations between our sample galaxies and whether or
not they have nearby, relatively massive (as defined in Sect.~2 as not
more than 3~mag fainter than the sample galaxy) companions, we
consider two related sets of numbers: those describing how many
galaxies in each of our nuclear and circumnuclear \ha\ morphology
categories has a close companion, and those describing the properties
of the subsets of sample galaxies that either have, or do not have, a
close companion. The result on the former, shown in
Table~\ref{overalltab} and already mentioned in Sect.~5.1, is that we
find no significant differences in the numbers of galaxies that have
close companions among our different categories of nuclear and
circumnuclear \ha\ morphology. This is surprising, because one might
have expected close companions and/or interactions to lead to enhanced
massive star formation in, and thus \ha\ emission from, the nucleus,
or nuclear rings (Knapen et al. 1995a; Buta \& Combes 1996; Heller \&
Shlosman 1996; Knapen et al. 2004b). In fact, the only hint at such a
general effect is the slightly lower fraction of galaxies with close
companions among those with no circumnuclear emission, but with two
out of nine galaxies causing the lower number in
Table~\ref{overalltab} we cannot call this deviation significant. We
must conclude that {\it neither} the nuclear, {\it nor} the
circumnuclear massive star formation morphology is at all affected by
the presence of close companions.

The results on the second test are summarised in
Table~\ref{companion_props}, which compares the properties of those
galaxies that have a close companion with those that do not. Columns~3
and 4 show that within our sample the galaxies with close companions
are further away, and smaller in angular diameter, than those without
close companions. This is related to the fact that the median absolute
magnitudes of the galaxies with close companions are brighter than
those of the isolated galaxies, in the median by 0.3~mag in $M_B$.  It
is plausible that the combination of these differences implies that
galaxies with close companions are brighter, and therefore more likely
to be included in our sample at higher distances.  Galaxies with a
close companion are, in the median, also of later morphological
$T$-type than those without a close companion. We performed
KS tests on these data sets, and found that
whereas the difference in distance between the two sub-samples is
significant ($P=0.029$), none of the other differences are ($D$,
$M_B$, $T$).

Yet even though our ``closely accompanied'' galaxies are brighter,
possibly because of enhanced star formation triggered by the
interaction, the effects on the morphology of the current massive star
formation in the nucleus and in the 2~kpc radius circumnuclear region
are negligible. Table~\ref{companion_props} confirms this by showing
the same data as tabulated before in Table~\ref{overalltab}, but now
analysed as percentages of the different \ha\ morphology classes
occurring in the sample galaxies with, and without, a close
companions. The percentages are remarkably similar, confirming our
overall conclusion that the presence of a close companion has no
influence on the morphology of massive star formation in and around
the centre. 

\subsection{Starburst and AGN activity}

\begin{table*}
\centering 
\begin{tabular}{lccccc}
\hline
 & Total & AGN & \hii/SB & non-AGN & non-AGN non-SB\\
\hline
\multicolumn{6}{c}{Nuclear emission}\\
Sample size & 72 & 31 & 5 & 41 & 36\\
Strong & 44 & 26 & 1 & 18  & 17 \\
Weak &   10 & 3  & --& 7   & 7 \\
None &   18 & 2  & 4 & 16  & 12 \\
\multicolumn{6}{c}{Circumnuclear emission}\\
Sample size & 73 & 31 & 5 & 42 & 37\\
Patchy & 39 & 13 & 5  & 26 & 21 \\
None &   9  & 5  & -- & 4  & 4 \\
Ring &   22 & 13 & -- & 9  & 9 \\
Diffuse & 3 & -- & -- & 3  & 3 \\
\hline
\end{tabular}
\caption{Nuclear and circumnuclear \ha\ emission for different
categories of nuclear activity, basically AGN (including LINER-type
activity) and starburst. Column~2: total sample.  Column~3: AGN, which
includes all galaxies classified in NED as either Seyfert or LINER.
Column~4: \hii/SB, all galaxies classified by NED as starburst or
\hii\ (excluding those whose classification also includes a LINER or
Seyfert which have been classed AGN).  Column~5: All galaxies not
classified as Seyfert or Liner. Column~6: as col.~5, but now also
excluding the galaxies of col.~4.  }
\label{agntab}
\end{table*}

We have scanned the NED for nuclear activity classifications of our
sample galaxies. As explained in K05, the information from NED may not
be as reliable as a uniformly conducted spectroscopic survey, although
K05 showed that this did not significantly alter their results. In
addition, our sample includes galaxies at larger distances than those
in K05. More galaxies may thus not have been checked for nuclear
activity in detail, and will hence, by default, have no nuclear
activity classifier in the NED. The results are tabulated in
Table~\ref{agntab}, which shows that of our 73 sample galaxies, 31
have been classified as AGN of some sort (we include Seyfert, LINER,
and ``AGN'' types in this category), with another 5 galaxies classed
as starburst using various nomenclature, including ``\hii''. 

The results, then, show two interesting features which are worth
discussing here. The first is the significantly higher fraction of AGN
which show strong nuclear \ha\ emission. Only a handful of galaxies,
in fact, show weak or no nuclear \ha\ emission. Any AGN should have
\ha\ emission, but the absence in the few cases where we report it can
be explained as a differential extinction effect. We use the
significantly redder $I$-band for the subtraction of the continuum
from the relatively bluer \ha\ line, which in a dust-extincted
environment will lead to surplus subtraction. This will lead to
artificially reduced \ha\ central peaks. Galaxies classified as
starbursts (or ``\hii'') but with ``none'' nuclear \ha\ emission may
well suffer from the same effects. Another possibility is that the
classification as AGN or starburst as given on the NED is in fact
erroneous, but checking that is outside the scope of this paper.

The second interesting feature from Table~\ref{agntab} is the enhanced
fraction of AGN among the nuclear ring host galaxies. Of the AGN, 42\%
have a ring, against 30\% of the complete sample. Conversely, 59\% of
the ring galaxies have an AGN, versus 42\% of the complete sample. K05
found a similar effect, but much more marked; there, almost all
nuclear rings were accompanied by AGN or starburst activity. 

The ring-AGN connection occurs partly because both rings and AGN are
most common among the early-type galaxies. Of our sample, 52 galaxies
(71\%) have $T\leq4$ (Sbc), but 21 out of our 22 rings (95\%) and 26
out of our 31 AGN (84\%) are hosted by galaxies of such early
types. This means that half of all $T\leq4$ galaxies are AGN, and
statistically 10.5 ring hosts will also host AGN. We concur with K05
in finding a higher number than this statistically expected one.  As
discussed in more detail by K05, the connection between rings and AGN
is most likely to be through the availability and recent inflow of
gas, which fuels both the massive star formation in the ring and
nuclear activity. In this interpretation, the nuclear rings are merely
a by-product of the gas inflow from the disk toward the very central
region (Shlosman 2005). A fraction of the inflowing gas moves further
inward, and is ultimately responsible for the AGN activity.

\subsection{Nuclear rings and their dynamical origin}

We identify 22 nuclear rings, three of which have not been reported in
the literature so far (those in NGC~473, NGC~5953, and
NGC~7716). These are star-forming nuclear rings, classified as such
purely on the basis of our \ha\ imaging. There may be a small number
of additional rings which are, for instance, too small, or too dusty
to be picked up by our imaging (the latter category might well include
NGC~1241, for which B\"oker et al. 1999 find a small nuclear ring in
Pa$\alpha$ which is absent from our \ha\ image). Of the 22 nuclear
rings, Table~\ref{sampletab} shows that four occur in galaxies classed
as unbarred (NGC~5953, NGC~6503, NGC~7217, and NGC~7742). In
addition, NGC~278, although classified as SX in the RC3, does not in
fact have a bar. Knapen et al. (2004b) made a detailed study of this
galaxy, and found from an analysis of near-IR and optical ground-based
images, as well as of {\it Hubble Space Telescope} ({\it HST})
imaging, that there is no evidence at all for the presence of a
bar. Radio 21~cm interferometry, however, showed that the outskirts of
the galaxy have a severely irregular and disturbed morphology and
kinematics in \hi, indicative of at least a minor merger event in the
recent history of the galaxy. Knapen et al. (2004b) postulate that
this minor merger lies at the origin of the massive star forming in
the nuclear ring, rather than a bar as is more conventional for
nuclear ring hosts.  Considering that the favoured formation scenario
of a nuclear ring is one where the ring forms in the vicinity of inner
Lindblad resonances set up by a non-axisymmetry in the gravitational
potential, due to, e.g., a bar or an interaction (see, e.g.,
Athanassoula 1992; Knapen et al. 1995b; Buta \& Combes 1996; Heller \&
Shlosman 1996; Knapen et al. 2004b), it is instructive to scrutinise
the four cases identified here of nuclear rings in supposedly
non-barred galaxies.

\begin{itemize}

\item NGC~5953 (classed as SA but interacting): this galaxy forms part
of a small interacting group of galaxies with NGC~5954 (visible in
Fig.~1) and UGC~9902. Hern{\' a}ndez-Toledo et al.(2003)
report the presence of a small ($\sim60$~pc) ``bar-like central
structure''. As in the case of NGC~278, it is plausible that this
interaction has led to the formation of the nuclear ring, and the
disturbed and complex \hi\ morphology and kinematics observed by
Chengalur et al.(1994) lends support to this suggestion.

\item NGC~6503 (classed as SA but inclined): this galaxy is rather
inclined ($i=70\deg$) and it is thus very unlikely that any bar can be
detected from optical or near-IR imaging -- we postulate from
statistical grounds only that this ring galaxy is most likely to be in
fact barred. The \hi\ velocity field is regular (Shostak, Willis, \&
Crane 1981; van Moorsel \& Wells 1985), but Bottema \& Gerritsen
(1997) find evidence for a separate kinematically distinct nuclear
component. We find no close companion.

\item NGC~7217 (classed as SA): this is a well-studied galaxy which in
fact has three rings (e.g., Combes et al. 2004 and references
therein). There is no evidence for a bar, but Merrifield \& Kuijken
(1994) reported the presence of the distinct counter-rotating disk
population. Although this is evidence for a past merger history, its
current gravitational effects are most probably limited. The velocity
field as seen from \hi\ mapping is regular (Verdes-Montenegro, Bosma
\& Athanassoula 1995). Buta et al. (1995) and Combes et al. (2004)
show that there is an oval distortion in this galaxy, which can
explain the occurrence of its three rings (nuclear, inner,
outer). This may well be an indicator of a historical, but now
destroyed, bar.

\item NGC~7742 (classed as SA): This is another well-known ring
galaxy, without any evidence for a bar or other asymmetry (Rix \&
Zaritsky 1995; Kornreich, Haynes, \& Lovelace 1998). No \hi\ velocity
field is available from the literature. De Zeeuw et al. (2002)
describe, on the basis of SAURON integral field spectroscopy, that the
gaseous ring counter-rotates with respect to the central stellar
component.

\end{itemize}

We thus summarise that of the four nuclear ring host galaxies
classified as SA in the RC3, one is interacting, and one is too
inclined to allow us to see any bar it may have. The remaining two
galaxies (NGC~7217 and NGC~7742) do not have a bar. NGC~7217 has an
oval distortion, which is able to maintain the nuclear ring, as shown
by simulations (Combes et al. 2004). It is possible that the nuclear
ring was formed during a recent bar episode, of which the oval
distortion is the remnant. This scenario is supported by the presence
of three rings in this galaxy, at the right resonant
locations. NGC~7742 has not been studied in nearly as much detail, has
no evidence for an oval, but does have counter-rotating gas and stars
in the nuclear region. 

NGC~278, although classified as SX in the RC3, does not have a bar,
but does show clear evidence for a recent minor merger (Knapen et
al. 2004b).  We can thus conclude that none of the galaxies reported
here challenges the standard scenario of nuclear ring formation to a
significant degree. In this scenario, gas and ensuing massive SF
accumulate near one or more inner Lindblad resonances, induced by an
asymmetry in the gravitational potential of the host galaxy, in turn
due to a bar or an interaction. Galaxies like NGC~278, NGC~7217, and
NGC~7742 which appear at first sight bar-less and isolated, remain
excellent test cases which warrant detailed studies.

In Paper~II, we will present a detailed discussion of the properties
of the 22 nuclear rings, and of how these properties relate to those
of their host galaxies and of their host bars, where those exist. We
will also discuss the identification and age determination of the
individual \hii\ region complexes or stellar clumps within the
rings. We will also compare the observed azimuthal age gradients, or
lack thereof, around the rings to dynamical models of ring formation.


\section{Conclusions}

We obtained a set of images in the $B$ and $I$ broad bands and in \ha
of a sample of 73 spiral galaxies. The data, obtained with the WHT and
mostly at a spatial resolution of below an arcsec, are presented here,
along with a classification of the morphology of the nuclear and
circumnuclear \ha\ emission. Most galaxies have relatively strongly
peaked nuclear \ha\ emission, and patchy emission from the 2\,kpc
radius circumnuclear region, the latter indicative of the presence of
individual \hi\ regions. We explore trends with host galaxy
parameters, and confirm that late-type galaxies have a patchy
circumnuclear appearance in \ha, and that nuclear rings occur
primarily in spiral types Sa-Sbc. We identify a number of previously
unknown nuclear rings. Although we report a number of cases where a
nuclear ring is hosted by an unbarred galaxy, we confirm that nuclear
rings are predominantly hosted by barred galaxies. Bars thus stimulate
nuclear rings, but do not influence the relative strength of the
nuclear \ha\ peak, nor other aspects of the circumnuclear \ha\
morphology. Even though our selection criteria led to an
over-abundance of galaxies with close massive companions, the presence
or absence of a close companion does not have {\it any} significant
influence on the relative strength of the nuclear \ha\ peak, nor on
the \ha\ morphology around the nucleus. A more detailed description of
the stellar populations and dynamical origin of the nuclear rings
described here will follow in Paper~II. The images described here will
be made available to the community.

\begin{acknowledgements}

JHK acknowledges support from the Leverhulme Trust in the form of a
Leverhulme Research Fellowship. We thank Dr. Koji Murakawa for
assistance during the 1999 observing run. LC thanks the University of
Hertfordshire for financial assistance. This research has been
partially supported by NASA/LTSA 5-13063, NASA/ATP NAG5-10823,
HST/AR-10284 and by NSF/AST 02-06251.The WHT is operated on the island
of La Palma by the Isaac Newton Group in the Spanish Observatorio del
Roque de los Muchachos of the Instituto de Astrof\'\i sica de
Canarias.  This research has made use of the NASA/IPAC Extragalactic
Database (NED) which is operated by the Jet Propulsion Laboratory,
California Institute of Technology, under contract with the National
Aeronautics and Space Administration.

\end{acknowledgements}

\label{lastpage}

\end{document}